\documentstyle[emulateapj]{article}

\newcommand{\sbb}{mag/$\sq\arcsec$}

\def\h2{H{\small II}}
\def\tol{{Tol\,1214--277}}
\begin{document}
\title{An imaging and spectroscopic study of the very metal-deficient 
blue compact dwarf galaxy 
Tol\ 1214--277\footnote{Based on observations obtained at the European Southern Observatory, Paranal, 
Chile (ESO Program 63.P-0003).}}

\author{Klaus J. Fricke}
\affil{Universit\"ats--Sternwarte, Geismarlandstra\ss e 11,
                 D--37083 G\"ottingen, Germany 
\\ Electronic mail: kfricke@uni-sw.gwdg.de}
\author{Yuri I. Izotov}
\affil{Main Astronomical Observatory, National Academy of Sciences of Ukraine,
Golosiiv, Kyiv 03680, Ukraine 
\\ Electronic mail: izotov@mao.kiev.ua}
\author{Polychronis Papaderos}
\affil{Universit\"ats--Sternwarte, Geismarlandstra\ss e 11,
                 D--37083 G\"ottingen, Germany 
\\ Electronic mail: papade@uni-sw.gwdg.de}
\author{Natalia G. Guseva}
\affil{Main Astronomical Observatory, National Academy of Sciences of Ukraine,
Golosiiv, Kyiv 03680, Ukraine 
\\ Electronic mail: guseva@mao.kiev.ua}
\and
\author{Trinh X. Thuan}
\affil{Astronomy Department, University of Virginia, Charlottesville, VA 22903
\\ Electronic mail: txt@virginia.edu}


\begin{abstract}

We present a spectrophotometric study based on 
VLT/FORS\,I observations of one of the most metal-deficient 
blue compact dwarf (BCD) galaxies known, \tol\ ($Z$$\sim$$Z$$_{\odot}$/25).
The data show that roughly half of the total luminosity of the BCD
originates from a bright and compact starburst region
located at the northeastern tip of a faint dwarf galaxy with cometary appearance.
The starburst has ignited less than 4 Myr ago and its emission is powered by several 
thousands O7V stars and $\sim$ 170 late-type nitrogen Wolf-Rayet stars located 
within a compact region with $\la$500 pc in diameter. 
For the first time in a BCD, a relatively strong 
[Fe\,V] $\lambda$4227 emission line is seen which together with 
intense He II $\lambda$4686 emission 
indicates the presence of a very hard radiation field in \tol. 
We argue that this extraordinarily hard radiation
originates from both Wolf--Rayet stars and 
radiative shocks in the starburst region.
The structural properties of the low-surface-brightness (LSB) component 
underlying the starburst have been investigated by means 
of surface photometry down to 28\ $B$ \sbb. 
We find that, for a surface brightness level fainter than $\sim 24.5$\ $B$\ \sbb,
an exponential fitting law provides an adequate approximation to its
radial intensity distribution. 
The broad-band colors in the outskirts of the LSB component of \tol\ 
are nearly constant and are consistent with an age below one Gyr.
This conclusion is supported by the comparison of the observed
spectral energy distribution (SED) of the LSB host with theoretical SEDs.
\end{abstract}

\section {Introduction \label{intro}}
Spectrophotometric studies of extremely metal-deficient
($Z<Z_{\odot}/20$) blue compact dwarf (BCD) galaxies give 
important insights into the properties of massive 
stars and the physical conditions in low-metallicity 
environments.
For example, recent spectroscopic studies of the most metal-deficient BCD 
known so far, I Zw 18, 
 unveiled a Wolf-Rayet (WR) stellar population 
(Izotov et al. 1997a, Legrand et al. 1997), which allows to  put
important constraints on evolutionary models for massive stars
with a very low heavy element abundance.
The analysis of such rare systems in the local Universe can 
also be of great benefit 
to studies of very high-redshift galaxies some of which may still be in their formation stages.
Further, the low metallicity and high excitation H II regions found in extremely 
metal-deficient BCDs allow to derive with high precision the primordial 
$^4$He abundance and hence one of the important 
cosmological parameters -- the baryon mass fraction of the Universe 
(e.g. Izotov et al. 1994, 1997c, 1999). 

Some recent observational studies indicate that galaxy formation is not a 
process occurring only at an early cosmological epoch but 
still happening at present, though at a very low level (Thuan et al. 1997, 
Papaderos et al. 1998, Izotov \& Thuan 1999, Kniazev et al. 2000). 
Izotov \& Thuan (1999) have proposed that a metallicity $<$1/20 $Z_{\odot}$
might be a useful indicator for identifying 
active dwarf galaxies in an unevolved stage. 
If so, the spectrophotometric analysis
of such nearby young galaxy candidates with age $\la$10$^9$ yr, 
in particular dwarf 
systems undergoing one of their first starbursts, is evidently of major
importance to observational cosmology.
On the other hand, some authors reach the conclusion 
that none of the very metal-deficient BCDs
at low redshift are young galaxies (e.g., Kunth \& \"Ostlin 2000, 
Legrand 2000, Legrand et al. 2000). The low-metallicity condition is 
fulfilled only in a tiny fraction ($\le$ 0.1 \%) 
of BCDs known to be gas-rich systems undergoing recurrent bursts of star-formation.
While debates concerning the age of extremely metal-deficient
BCDs continue it seems helpful to proceed with detailed studies of all these
galaxies identified so far in the local Universe.

We focus here on one of these rare candidates, the BCD Tol 1214--277,
the very low metallicity of 
which has been established by earlier spectroscopic work 
(Campbell et al. 1986; Pagel et al. 1992; Masegosa, Moles \& 
Campos-Aguilar 1994). 
Thuan \& Izotov (1997) have derived the distance to \tol\ of $D$ = 
103.9 Mpc adopting the observed radial velocity $v$ = 7795 km s$^{-1}$ 
and assuming the Hubble constant $H_0$ = 75 km s$^{-1}$ Mpc$^{-1}$.
The presence of very massive stars in this BCD is indicated by strong 
nebular lines (Pagel et al. 1992) and a UV stellar N\,V $\lambda$1240 
line with a P\,Cygni profile (Thuan \& Izotov 1997).
The latter properties along with the fact that \tol\ is
the lowest metallicity BCD with detected Ly$\alpha$ emission 
(Thuan \& Izotov 1997) make it not only a probable young BCD candidate but also
allow the study of the Ly$\alpha$ radiation escaping from the starburst.

Telles et al. (1997) described \tol\ as a single-knot source with extensions 
along the north-south direction. Surface photometry studies by the same
authors revealed a compact ($\la 3$\arcsec) high-intensity
 core on top of a much fainter component with an   
exponential profile of angular scale length of 3\farcs4.
However, little is known on the structural properties and colors of that
diffuse stellar host underlying the starburst component.
The determination of these properties will be crucial for assessing the evolutionary 
state of Tol 1214--277.

In the following, we investigate the photometric structure and 
spectral properties of \tol\ with deep VLT data.
In Sect. 2 we describe the data, and in Sects. 3 and 4 we discuss
the results obtained, respectively, from broad-band imaging and long-slit 
spectroscopy. We discuss the age of \tol\ in Sect. 5 and summarize 
our results in Sect. 6.
%


\begin{figure*}[tbh]
\figurenum{1}
\plotfiddle{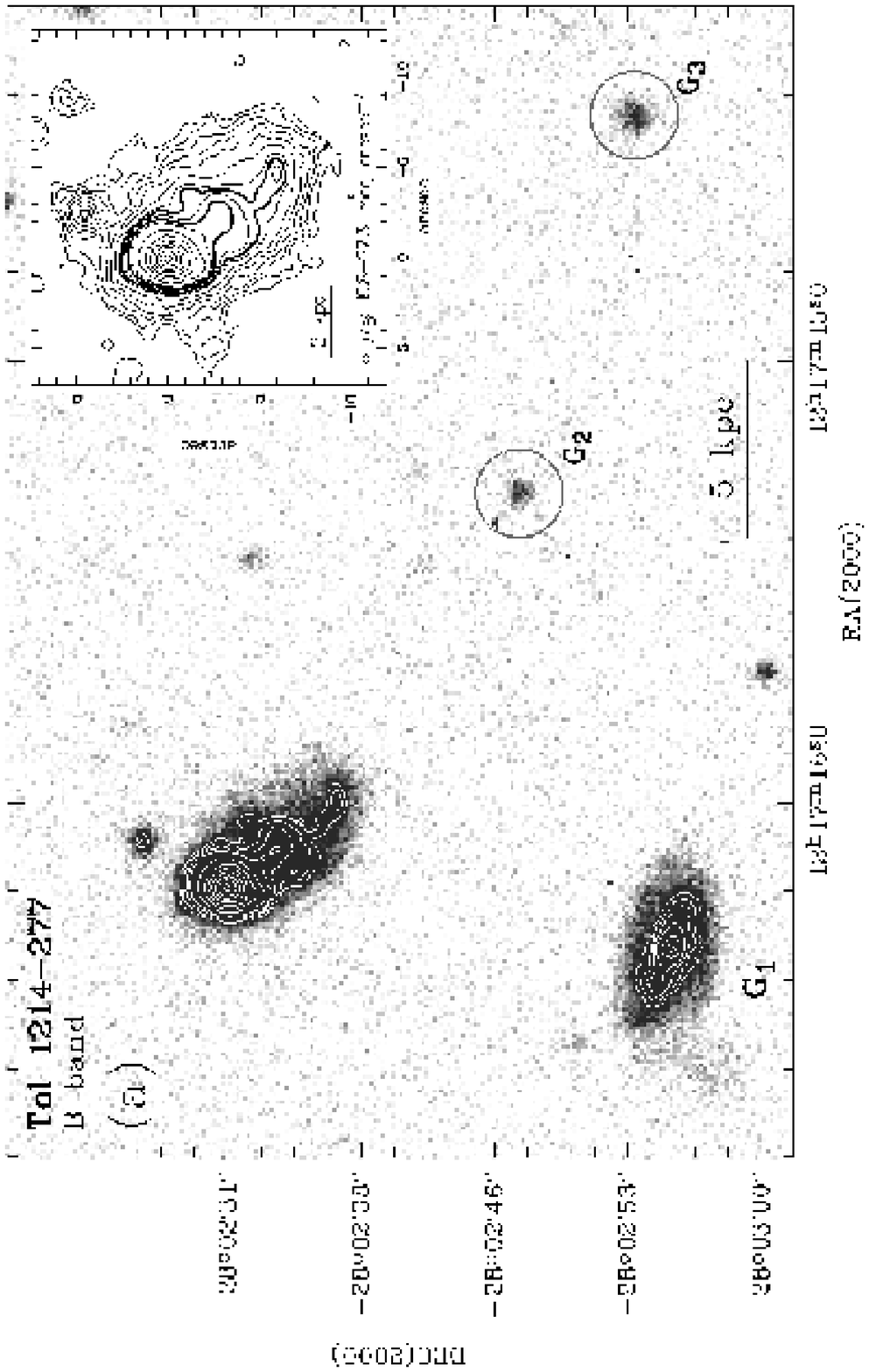}{0.cm}{270.}{50.}{50.}{-300.}{30.}
\plotfiddle{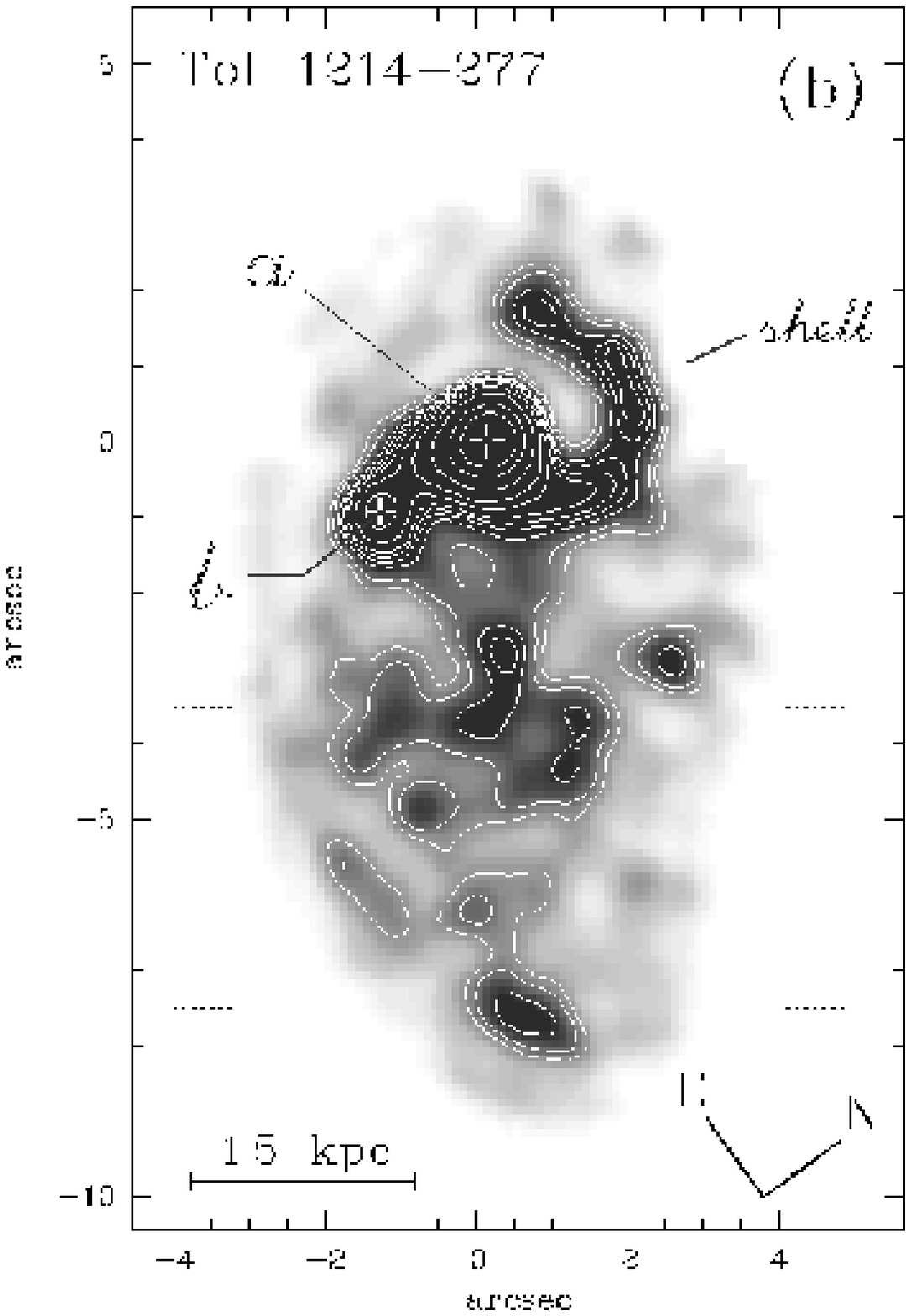}{0.cm}{0.}{37.}{37.}{50.}{-245.}
\vspace{7.cm}
\figcaption{\label{f1} (a) $B$ band exposure of the iI,C--BCD \tol\ obtained with the VLT/FORS. North is up and east is to the left.
The bright star-forming complex at the northeastern extreme end of the elongated 
LSB host is visible.
The images show some clumpiness along the irregular pattern of the LSB body with
two local luminosity maxima 3\farcs 5 (1.76 kpc) and 7\farcs 5 (3.8 kpc) southwest of  
the starburst nucleus.
The inset to the upper right shows a contour map of the BCD.
Contours correspond to intensity levels of 23.5 and 24 $B$\ \sbb\ (thick lines) 
and from 19.8 to 27.3 $B$\ \sbb\ in steps of 0.5 mag (thin lines).
(b) $R$ band exposure of \tol\ deconvolved with 30 iterations of the 
Lucy algorithm (Lucy 1974). The angular resolution is improved to $\approx$0\farcs 46. 
The local luminosity maxima seen in the original exposures 
along the main body of the BCD at surface brightness levels 
of $\approx$ 23.5 to 24.0 $B$\ \sbb\ (dotted horizontal lines)
are resolved into an assembly of compact (diameter $\la$0\farcs 5)
sources of medium surface brightness.
The starburst component splits into 2 high surface brightness knots with an 
angular separation of $\approx 2$\arcsec. 
Roughly 1 kpc northwards of the most luminous knot {\it a},
the deconvolution reveals a curved feature ({\it shell})
probably a starburst-driven supershell.}
\end{figure*}

\begin{figure*}[tbh]
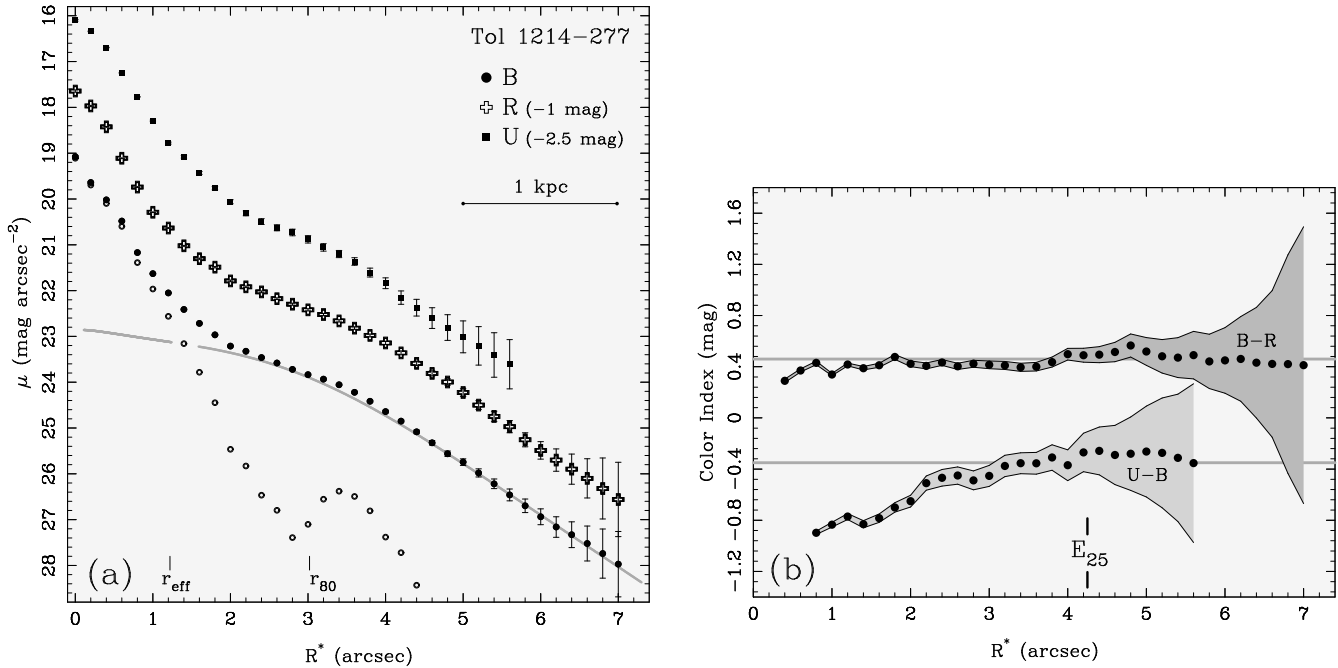

\figurenum{2}
\plotfiddle{Fig2a.ps}{0.cm}{270.}{50.}{50.}{-290.}{40.}
\plotfiddle{Fig2b.ps}{0.cm}{270.}{50.}{50.}{-30.}{64.}
\vspace{8.cm}
\figcaption{\label{f2} (a) Surface brightness profiles (SBPs) of 
\tol\ derived from VLT $U$, $B$ and $R$ images. The SBPs in $U$ and $R$ 
are shifted vertically by --2.5 mag and --1 mag, respectively, for 
the sake of clarity.
The 2$\sigma$ photometric uncertainty for each point along the SBPs is shown by
vertical bars.
The radii $r_{\rm eff}$ and $r_{80}$ which encircle, respectively, 50\% and 80\% 
of the total $B$ band emission are indicated. 
It is evident that the exponential fit to the LSB component 
for surface brightness levels $\mu_{\rm B}$ $\ga$ 24.5 \sbb\ cannot be 
extrapolated inwards without producing a luminosity excess for photometric radii 
$r_{\rm eff}$$\la$$R^*$$\!<$4\arcsec. We have therefore modelled the intensity 
distribution of the LSB host adopting a modified exponential fitting law 
(Eq. \ref{eq:p96a}) which allows for a central flattening (thick curve).
The surface brightness distribution of the starburst (small open circles) 
is inferred from the luminosity in excess to the fit of the LSB host.
(b) Radially averaged ($U-B$) and ($B-R$) color profiles of \tol\ 
computed by subtraction of SBPs displayed in panel (a).
The average colors of the LSB host, as obtained by subtraction of 
exponential models fitted to its intensity distribution in different bands 
(cf. panel (a), thick curve), of --0.35 mag and 
+0.45 mag for ($U-B$) and ($B-R$), respectively, are shown by 
horizontal lines. The isophotal radius E$_{25}$\ 
of the LSB host at 25 $B$\ \sbb\ is indicated.}
\end{figure*}

%
\section{Observations and data reduction \label{obs}}
\subsection{Photometric data}
Images of \tol\ in the broad-band filters Bessell $U$, $B$, $R$ were obtained 
with the {\sf Fo}cal {\sf R}educer and low-dispersion {\sf S}pectrograph
(FORS\,1; see Moehler et al. 1995) attached to the VLT\ UT1.
The exposures were acquired under photometric conditions on May, 17th 1999 
during a 6 night run allocated to guaranteed time observations (GTO)
at an airmass ranging between 1.48 and 1.64.
The seeing was between 0\farcs 7 and 0\farcs 9. FORS was operating in the 
standard imaging mode yielding a final focal ratio of 3.13 and an instrumental 
scale of 0\farcs 2 pix$^{-1}$.

Photometric zero-points and color-dependent calibration terms 
were derived from exposures of the standard-star field Mark\ A 
(Landolt 1992) taken each night during the GTO run at an 
airmass$\approx$1. 
Airmass-dependent calibration terms were obtained using standard 
extinction curves and found to agree to a level better than 10\% 
with those derived during the commissioning phase\,II of FORS.
The photometric precision is estimated to be $\sim 0.1$ mag.
Reduction has been accomplished in the standard way using the ESO-MIDAS 
software package. 
\subsection{Spectroscopic data}
Spectroscopic data for Tol 1214--277 were taken with FORS\,1
on May, 12th 1999 at an airmass 1.7 and with a seeing 
between 0\farcs7 -- 0\farcs9 FWHM. 
A 1\arcsec\ $\times$ 180\arcsec\ slit was used in conjunction 
with a grism GRIS$\_$300V and a GG375 second-order blocking filter.
This yields a spatial resolution along the slit of 0\farcs2 pixel$^{-1}$,
a scale perpendicular to the slit of 3 \AA\ pixel$^{-1}$, a spectral
coverage of 3600 -- 7500 \AA, and a spectral resolution of $\sim$ 10 \AA\ (FWHM).
The total exposure time of 1650 seconds has allowed to reach
a signal-to-noise ratio S/N $\ga$ 50 in the continuum of the 
bright central part of the BCD and was broken up into two 
subexposures, 990 and 660 seconds, to allow for an efficient
cosmic-ray removal. 
The slit was oriented in the position angle P.A. = 39$^{\circ}$ to enable 
a simultaneous study of the starburst knot and the faint underlying host 
along its major axis. A spectrum of a He-Ne-Ar comparison lamp was obtained 
for wavelength calibration. 

Since no spectrophotometric standard star was observed,
calibration was done using flux-calibrated 
spectra of \tol\ obtained previously with the 2.1m KPNO 
telescope\footnote{Kitt Peak National Observatory (KPNO) is operated by 
the Association of Universities for 
Research in Astronomy (AURA), Inc. under cooperative agreement with the 
National Science Foundation.}. 
These were taken on April 2, 1998 at an airmass 2.0.
The total exposure time was 3600 seconds, split into 3 subexposures, 
1200 seconds each. A 2\arcsec\ $\times$ 180\arcsec\ slit was used along 
with grating No. 9 and a GG375 second-order blocking filter.
This yields a spatial resolution along the slit of 0\farcs69 pixel$^{-1}$,
a scale perpendicular to the slit of 2.7 \AA\ pixel$^{-1}$, a spectral
range 3600 -- 7500 \AA, and a spectral resolution of $\sim$ 7 \AA\ (FWHM). The 
spectrophotometric standard star Feige 34 was observed for flux calibration. 
Data reduction of the 2.1m telescope and VLT spectroscopic 
observations was carried out using the IRAF software package. 
This included bias subtraction, cosmic-ray removal,
flat-field correction, wavelength calibration and night-sky 
background subtraction.
%

\section{Photometric analysis \label{phot}}
\subsection{Morphology of the starburst component of \tol\ \label{starburst}}
%
%

\begin{deluxetable}{lccccccccccc}\label{Fricke.tab1}
\tablenum{1}
\tablecolumns{12}
\tablewidth{0pt}
\tablecaption{Photometric properties of the starburst and LSB components of 
\tol\tablenotemark{a} \label{Tab1}}
\tablehead{
\colhead{Band} & \colhead{$t$}& \colhead{$\mu _{\rm E,0}$} 
& \colhead{$\alpha$} & \colhead{P$_{25}$}  &\colhead{$m_{\rm P_{25}}$} 
& \colhead{E$_{25}$} & \colhead{$m_{\rm E_{25}}$} & \colhead{$m_{\rm LSB}^6$} 
& \colhead{$m_{\rm SBP}$} & \colhead{$m_{\rm Polyg}$} 
& \colhead{$r_{\rm eff}$,$r_{80}$} \nl 
\colhead{}& \colhead{min} & \colhead{mag arcsec$^{-2}$} &\colhead{pc}
& \colhead{kpc}&\colhead{mag}&\colhead{kpc}&  \colhead{mag}& \colhead{mag}
& \colhead{mag}&\colhead{mag}&\colhead{kpc} \nl
\colhead{(1)}&\colhead{(2)}&\colhead{(3)}&\colhead{(4)}&\colhead{(5)}
&\colhead{(6)}&\colhead{(7)}&\colhead{(8)}&\colhead{(9)}&\colhead{(10)}
&\colhead{(11)}&\colhead{(12)}
}
\startdata
 $U$ & 10   & 19.69$\pm$0.4  & 468$\pm$73  & 1.18  & 18.28  & 2.27  & 18.99  & 18.89 & 17.75$\pm$0.12 & 17.72 & 0.50,1.22 \nl 
 $B$ & 6    & 20.11$\pm$0.2  & 483$\pm$16  & 0.96  & 19.20  & 2.14  & 19.39  & 19.24 & 18.42$\pm$0.02 & 18.42 & 0.61,1.52\nl 
 $R$ & 4    & 19.64$\pm$0.2  & 487$\pm$19  & 0.98  & 18.72  & 2.38  & 18.84  & 18.78 & 17.95$\pm$0.03 & 17.94 & 0.59,1.51\nl 
\enddata
\tablenotetext{a}{The values above have not been corrected for extinction.
A distance of 103.9 Mpc has been adopted throughout. 
The central surface brightness $\mu_{\rm E,0}$ listed in col.\ 3 
is obtained by extrapolation of the exponential 
slope observed in the outskirts of \tol\ for radii $R^*$$\geq$4\arcsec\
to $R^*$=0\arcsec. A more appropriate model for the intensity 
distribution of the LSB host of the form of Eq. 2} 
(Fig. \ref{f2}a; thick curve) implies for all bands a 
central surface brightness 2.74 mag fainter than the listed
$\mu_{\rm E,0}$.
\end{deluxetable}

The VLT $B$ image (Fig.\ 1a) reveals an elongated 
LSB component extending $\sim$ 11\arcsec\ (5.5 kpc) to the southwestern
direction of the unresolved high-surface-brightness (HSB) starburst nucleus.
Such properties place \tol\ in the rare iI,C ``cometary'' BCD class of the 
Loose \& Thuan (1985) morphological classification. As for
 the three nearby extended sources 
indicated in Fig.\ 1a, the previously non-catalogued source G$_1$ is
an inclined SBb galaxy most likely unrelated to the BCD. 
The nature of two faint sources located close to the southwestern 
prolongation of the LSB host of \tol\ labelled G$_2$ ($m_B$=23.5$\pm$0.2) 
and G$_3$ ($m_B$=22.8$\pm$0.2) is not known. 
In view of the recent findings by van Zee et al. (1998ab) of
elongated gaseous reservoirs pointing in some cases over kpc scales towards the direction
of the major axis of the BCD, it may be worth checking
with follow-up spectroscopy and radio H\,I interferometry
a possible physical connection of the sources G$_2$ and G$_3$ with \tol.

The deconvolved image of \tol\ in the R band (Fig.\ \ref{f1}b) shows that 
the starburst component contains at least two star-forming regions with an 
angular separation of $\sim2$\arcsec\ ($\sim 1$ kpc). 
Roughly 90\% of the total starburst luminosity in the $R$ band originates from
knot $a$ (diameter $\la$1\arcsec; 0.5 kpc). As suggested by its 
slight extension to the south, another unresolved knot may be embedded therein. 
The adjacent source labelled $b$ contributes less than 5\% 
of the starburst emission in the $R$ band.

A prominent curved structure can be seen in the deconvolved image of
Tol 1214--277,
at a projected distance of $\sim 1.1$ kpc northwards of 
starburst region $a$. This is likely 
a starburst-driven supershell. 
Circumstantial evidence for a partial disruption of the absorbing 
H I halo (as illustrated for example by case $b$ in
 Fig.\ 8 of Tenorio-Tagle et al. 1999)
of \tol\ as a consequence of the energetic output of the burst, comes 
from the strong Ly$\alpha$ emission with an equivalent width of 
70${\rm \AA}$ (Thuan \& Izotov 1997), the highest value observed 
yet in a BCD.
%
\subsection{Surface photometry \label{lsb}}
%

Surface brightness profiles (SBPs) of \tol\ (Fig.\ \ref{f2}a) 
were derived employing methods described in Papaderos et al. 
(1996b; hereafter P96) and interpreted in terms of a simplified 
starburst/LSB decomposition scheme. 
%
The photometric uncertainty was assigned to each point along the SBPs
assuming that Poisson statistics apply and taking into account inaccuracies 
in the determination of the sky background level (cf. P96).

The SBPs of \tol\ display two intensity 
regimes. First, the steeply rising intensity core for radii $<\!2$\arcsec\ 
($\mu_B\!<\!23$\ \sbb) can be attributed to the bright starburst region 
at the northeastern tip of the LSB host. 
Second, an exponential intensity regime is visible 
for $\mu\ga24.5$ $B$ \sbb\ down to 28\ $B$\ \sbb, i.e. beyond the photometric radius
$R^*>4$\arcsec\ which encircles $\sim 90$\% of the total $B$ band emission of the BCD. 
From Fig.\ \ref{f2}a, it is evident, however, that an 
inwards extrapolation of the exponential slope characterizing the 
outskirts of the galaxy leads for radii $r_{\rm eff}\la R^* \la 4$\arcsec\ 
to an intensity higher than that observed.
Therefore, a proper approximation to the intensity distribution of 
the LSB host seems possible only when a flattening of the exponential 
fitting law inwards of $\sim$ 3 scale lengths is postulated. 
Such an exponential profile levelling off for small radii has been 
frequently found in low-luminosity dwarf ellipticals (type--V profile; 
cf. Binggeli \& Cameron 1991), dwarf irregulars (Patterson \& Thuan 1996, 
Makarova 1999, van Zee 2000) and in few BCDs (cf. Vennik et al. 1996, 
P96, Telles et al. 1997, Papaderos et al. 1999).
A S\'ersic fitting law (S\'ersic 1968) of the form
\begin{equation}
I(R^*) = I_0\,\exp\left(-\frac{R^*}{\alpha}\right)^{\eta}
\label{eq:sersic} 
\end{equation}
where $I(R^*)$ is the observed intensity at the photometric
radius $R^*$ and $\alpha$ the angular scale length.
An exponent ${\eta}\sim 2.2$ and a scale length of $\sim$ 3\farcs 5 
reproduces well the central intensity flattening, but provides only a 
moderately good fit to the outer exponential slope for $R^*$$>$4\arcsec.

\begin{figure*}
\figurenum{3}
\plotfiddle{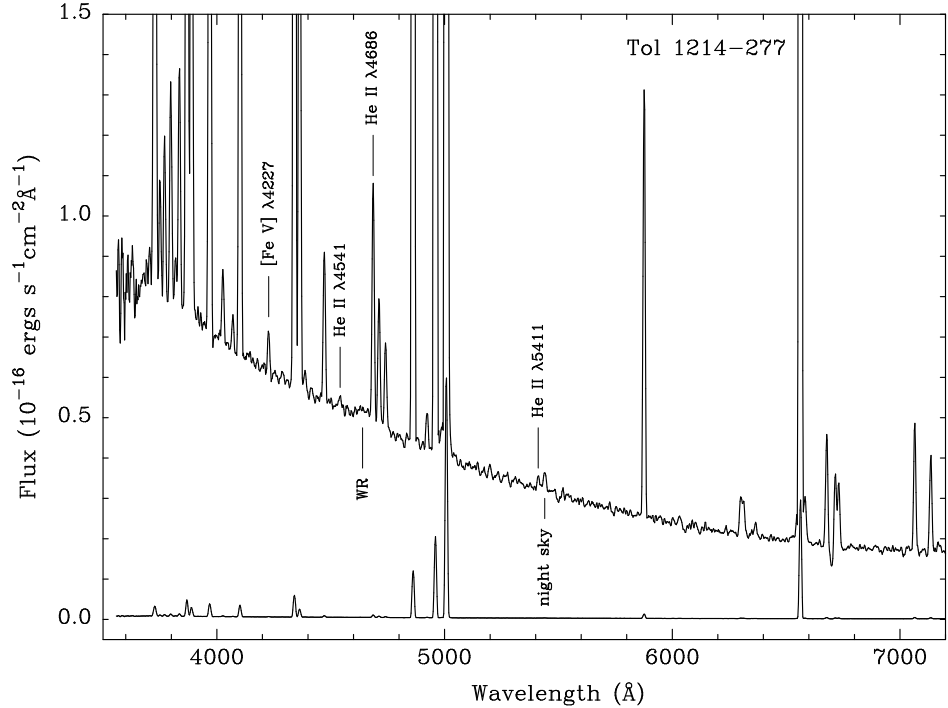}{0.cm}{270.}{60.}{60.}{-260.}{70.}
\vspace{9.cm}
\figcaption{\label{f3} The spectrum of the brightest part of Tol 1214--277.
The Wolf-Rayet bump and emission lines of high-ionization species are
indicated. 
The lower spectrum is the observed spectrum downscaled by a factor of 
100.}
\end{figure*}

%
%

\begin{deluxetable}{lccccc}
\tablenum{2}
\tablecolumns{2}
\tablewidth{200pt}
\tablecaption{Emission line intensities in the brightest knot of Tol 1214--277\label{Fricke.tab2}}
\tablehead{
\colhead{Ion}&\colhead{$I$($\lambda$)/$I$(H$\beta$)}}
\startdata
 3727\ [O II]        &0.2438$\pm$0.0027\nl
 3750\ H12           &0.0213$\pm$0.0020\nl
 3770\ H11           &0.0322$\pm$0.0021\nl
 3798\ H10           &0.0480$\pm$0.0022\nl
 3820\ He I          &0.0094$\pm$0.0014\nl
 3835\ H9            &0.0517$\pm$0.0022\nl
 3868\ [Ne III]      &0.3429$\pm$0.0030\nl
 3889\ He I + H8     &0.1906$\pm$0.0027\nl
 3968\ [Ne III] + H7 &0.2859$\pm$0.0030\nl
 4026\ He I          &0.0158$\pm$0.0015\nl
 4069\ [S II]        &0.0071$\pm$0.0012\nl
 4101\ H$\delta$     &0.2536$\pm$0.0028\nl
 4227\ [Fe V]        &0.0095$\pm$0.0013\nl
 4340\ H$\gamma$     &0.4793$\pm$0.0037\nl
 4363\ [O III]       &0.1731$\pm$0.0021\nl
 4388\ He I          &0.0047$\pm$0.0013\nl
 4471\ He I          &0.0352$\pm$0.0015\nl
 4541\ He II         &0.0023$\pm$0.0013\nl
 4650\ WR bump       &0.0161$\pm$0.0015\nl
 4686\ He II         &0.0527$\pm$0.0015\nl
 4711\ [Ar IV] + He I&0.0275$\pm$0.0014\nl
 4740\ [Ar IV]       &0.0185$\pm$0.0013\nl
 4861\ H$\beta$      &1.0000$\pm$0.0053\nl
 4922\ He I          &0.0086$\pm$0.0013\nl
 4959\ [O III]       &1.7396$\pm$0.0083\nl
 5007\ [O III]       &5.0926$\pm$0.0205\nl
 5200\ [N I]         &0.0027$\pm$0.0011\nl
 5411\ He II         &0.0036$\pm$0.0010\nl
 5876\ He I          &0.0935$\pm$0.0015\nl
 6300\ [O I]         &0.0099$\pm$0.0012\nl
 6312\ [S III]       &0.0059$\pm$0.0010\nl
 6363\ [O I]         &0.0033$\pm$0.0009\nl
 6563\ H$\alpha$     &2.5700$\pm$0.0120\nl
 6583\ [N II]        &0.0088$\pm$0.0021\nl
 6678\ He I          &0.0248$\pm$0.0009\nl
 6717\ [S II]        &0.0169$\pm$0.0010\nl
 6731\ [S II]        &0.0137$\pm$0.0010\nl
 7065\ He I          &0.0273$\pm$0.0009\nl
 7135\ [Ar III]      &0.0221$\pm$0.0009\nl \nl
 $C$(H$\beta$) dex    &\multicolumn {1}{c}{0.0}\nl
 $F$(H$\beta$)\tablenotemark{a} &\multicolumn {1}{c}{ 1.40$\pm$0.02}\nl
 $EW$(H$\beta$)\ \AA &\multicolumn {1}{c}{324$\pm$1}\nl
\enddata
\tablenotetext{a}{in units of 10$^{-14}$ ergs\ s$^{-1}$cm$^{-2}$.}
\end{deluxetable}

%
\begin{deluxetable}{lc}
\tablenum{3}
\tablecolumns{2}
\tablewidth{200pt}
\tablecaption{Element abundances in Tol 1214--277
\label{Fricke.tab3}}
\tablehead{
\colhead{Parameter}&\colhead{Value}}
\startdata
$T_e$(O III)(K)                     &20040$\pm$160 \nl
$T_e$(O II)(K)                      &15700$\pm$110 \nl
$T_e$(S III)(K)                     &18300$\pm$130 \nl
$N_e$(S II)(cm$^{-3}$)              &   210$\pm$160\nl
$N_e$(He II)(cm$^{-3}$)             &   150$\pm$90 \nl
$\tau$($\lambda$3889)               &       0.0    \nl \nl
O$^+$/H$^+$($\times$10$^5$)         &0.189$\pm$0.004\nl
O$^{+2}$/H$^+$($\times$10$^5$)      &2.929$\pm$0.052\nl
O$^{+3}$/H$^+$($\times$10$^5$)      &0.200$\pm$0.010\nl
O/H($\times$10$^5$)                 &3.317$\pm$0.053\nl
12 + log(O/H)                       &7.521$\pm$0.007\nl \nl
N$^{+}$/H$^+$($\times$10$^7$)       &0.602$\pm$0.013\nl
ICF(N)\tablenotemark{a}                              &17.6\,~~~~~~~~~~\nl
log(N/O)                            &--1.496$\pm$0.016~~\nl \nl
Ne$^{+2}$/H$^+$($\times$10$^5$)     &0.397$\pm$0.008\nl
ICF(Ne)\tablenotemark{a}                             &1.13\,~~~~~~~~~~\nl
log(Ne/O)                           &--0.868$\pm$0.013~~\nl \nl
S$^+$/H$^+$($\times$10$^7$)         &0.286$\pm$0.014\nl
S$^{+2}$/H$^+$($\times$10$^7$)      &1.728$\pm$0.293\nl
ICF(S)\tablenotemark{a}                              &3.78\,~~~~~~~~~~\nl
log(S/O)                            &--1.640$\pm$0.033~~\nl \nl
Ar$^{+2}$/H$^+$($\times$10$^7$)     &0.580$\pm$0.025\nl
Ar$^{+3}$/H$^+$($\times$10$^7$)     &1.448$\pm$0.105\nl
ICF(Ar)\tablenotemark{a}                             &1.01\,~~~~~~~~~~\nl
log(Ar/O)                           &--2.212$\pm$0.024~~\nl \nl
He$^+$/H$^+$($\lambda$4471)         &0.0732$\pm$0.0030\nl
He$^+$/H$^+$($\lambda$5876)         &0.0747$\pm$0.0013\nl
He$^+$/H$^+$($\lambda$6678)         &0.0745$\pm$0.0028\nl
He$^+$/H$^+$(weighted mean)         &0.0745$\pm$0.0011\nl
He$^{+2}$/H$^+$($\lambda$4686)      &0.0049$\pm$0.0001\nl
He/H                                &0.0793$\pm$0.0011\nl
$Y$                                 &0.2407$\pm$0.0034\nl
\enddata
\tablenotetext{a}{ICF is the ionization correction factor for unseen
stages of ionization. The expressions for ICFs are adopted from ITL94.}
\end{deluxetable}

Another empirical expression well suited for fitting an exponential 
distribution which is truncated inwards was proposed by P96 as
\begin{equation}
I(R^*) = I_0\,\exp\left( -\frac{R^*}{\alpha}\right)
\big[1-q\,\exp(-P_3(R^*))\big]
\label{eq:p96a} 
\end{equation}
where $P_3(R^*)$ is
\begin{equation}
P_3(R^*) = \left(\frac{R^*}{b\,\alpha}\right)^3+\left(\frac{R^*}{\alpha}\,\frac{1-q}{q}\right).
\label{eq:p96b} 
\end{equation}
The intensity distribution given by Eq. \ref{eq:p96a} depends near the center
on the relative central intensity depression
$q=\Delta I/I_0$, with $I_0$ being the central intensity of a pure exponential
law, and $b \alpha$ the cutoff--radius where the central flattening occurs.
By fitting the latter distribution to the SBP of \tol\ 
(Fig.\ \ref{f2}a; thick curve) we obtain a parameter set 
($b$,$q$)=(3.3,0.92) implying that an intensity depression 
occurs already within 3.3 exponential scale lengths and leads 
to a central intensity of $\sim$10\% of that predicted 
by an inwards extrapolation of the outer exponential slope.

As may be seen from the same figure, subtraction of 
our fitting model from the observed SBP (open circles 
in Fig. \ref{f2}a) allows to disentangle the brightness 
distribution of the starburst knot from that of the secondary 
faint assembly of sources $\sim$3\farcs 5 southwest of the 
starburst nucleus, seen in direct and deconvolved images 
(Fig.\ \ref{f1}a,b). 

Table\ 1 summarizes the derived photometric quantities. Cols.\ 3\&4 give, 
respectively, the central surface brightness $\mu_{\rm E,0}$ and scale 
length $\alpha$ of the LSB host as obtained from linear fits to the SBPs 
for $R^*\geq 4$\arcsec\ and weighted by the photometric uncertainty of 
each point. 
Cols. 5 though 9 list quantities obtained from profile decomposition 
whereby the intensity distribution of the LSB host was modelled by the 
modified exponential distribution Eq. \ref{eq:p96a}.
Cols.\ 5 and 7 give the radial extents P$_{25}$ and E$_{25}$
of the starburst and LSB components respectively, 
both determined at a surface brightness level 25\ \sbb.
The apparent magnitude of either component within P$_{25}$ and E$_{25}$
are listed in cols.\ 6 and 8, respectively. Col.\ 9 gives the apparent 
magnitude of the LSB component within a photometric radius of 6\arcsec.
Cols.\ 10 and 11 list the apparent magnitude as derived from integration of 
each SBP out to the last measured point and the total magnitude 
in each band inferred within a polygonal aperture. 
Col.\ 12 gives the effective radius $r_{\rm eff}$ and the radius 
$r_{80}$, which encircle 50\% and 80\% of the galaxy's total flux, 
respectively.

Profile integration implies that the starburst contributes nearly 
half of the $B$ band emission of \tol\ within its isophotal radius E$_{25}$ 
($\approx$4\farcs 6) while the remaining half is due to the underlying 
LSB dwarf for which Eq. \ref{eq:p96a} yields an absolute B magnitude of
--15.8 mag within its Holmberg radius.

The fractional luminosity contribution of the starburst for 
\tol\ compares well with the average value inferred for BCDs by 
Papaderos et al. (1996a) and Salzer \& Norton (1998).
This is also the case for the exponential scale length 
$\alpha\!\approx\!480$ pc for \tol\ which is in the range 
of those obtained for the LSB hosts in other extremely 
metal-deficient BCDs, such as SBS\ 0335--052\,E 
($\alpha$=458$\pm$8 pc; Thuan et al. 1997, Papaderos et al. 1998) 
and SBS\ 1415+437 ($\alpha$$\sim$ 300 pc; Thuan, Izotov \& Foltz 1999).
Such values are larger by factors 2--3 than the typical scale lengths of 
the LSB hosts of ultra-compact BCDs, as for instance Mkn\ 36 (130 pc; P96), 
Mkn\ 487 (190 pc; P96) and Pox\ 186 (180 pc; Doublier et al. 2000).
This implies that, despite its faintness and low angular extent, \tol\ 
is not an ultra-compact BCD.

%
\subsection{Color distribution}
%
Figure\ 2b shows that over the entire body of the galaxy, covering a 
range of $\sim 9$ mag in surface brightness, the ($B-R$) color 
remains nearly constant at (+0.44$\pm$0.05) mag.
By contrast, the ($U-B$) index is very blue ($\la$--0.8 mag) for radii 
$\la r_{\rm eff}$, increasing then steadily with a gradient of 
(0.33$\pm$0.04) mag\ kpc$^{-1}$ out to $R^*$$\sim$4\arcsec\,
beyond which it varies between the extremes of --0.4 mag and --0.26 mag. 
A more reliable interval for the ($U-B$) color of the LSB host is 
between --0.45 mag and --0.35 mag, the first value being the color   
at the radius $r_{80}$, where the contribution of the starburst has 
dropped below 5\% of the total luminosity and the second value resulting from
direct subtraction of the modelled intensity distributions of the LSB host 
in $U$ and $B$ (cf. Fig. \ref{f2}b).
Note that from $r_{\rm eff}$ to $r_{80}$ the gradual increase of the ($U-B$) 
index up to $\sim$--0.45 mag reflects the decrease of the starburst luminosity
while the ($B-R$) index remains roughly constant, presumably, as a result 
of the combined effects of stellar and gaseous emission.

As we shall show in Sect.\ 5 gaseous emission does not dominate the light
of \tol\ along its southwestern extension, therefore the broad band 
color indices in the outskirts of the LSB host for radii 
$\ga$4\arcsec\ (Fig. \ref{f2}b) can be directly used to 
estimate its age.

\section{Spectroscopic analysis \label{spec}}
\subsection{Element abundances}
\subsubsection{Emission-line intensities}

The spectrum of the brightest knot of Tol 1214--277 extracted within an
aperture of 2\arcsec\ $\times$ 1\arcsec\ is shown in Fig. \ref{f3}. 
It is dominated by very strong emission lines, reflecting the 
ongoing star formation activity. 
Remarkable spectral features are the nebular He II $\lambda$4686,
$\lambda$4541, $\lambda$5411 and [Fe V] $\lambda$4227 emission lines
suggesting a very hard stellar radiation field in the BCD.
Furthermore, a broad stellar bump at $\lambda$4650 is
detected indicating the presence of WR stars. 
Intensities of nebular lines have been measured by fitting Gaussians to the 
line profiles, while the intensity of the broad stellar bump has 
been derived by measuring its integral excess emission 
in the wavelength range $\lambda$4600 -- 4700 after 
subtraction of nebular emission lines. The errors of line intensities and 
equivalent widths include the errors
in placement of continuum and those in the Gaussian fitting. These errors
have been propagated in calculations of element abundances and the numbers
of O and WR stars.
The emission line intensities together with the equivalent width 
$EW$(H$\beta$) and absolute flux $F$(H$\beta$) of the
H$\beta$ emission line are listed in Table \ref{Fricke.tab2}. 
Because of the low observed H$\alpha$-to-H$\beta$ intensity ratio
we adopted an extinction coefficient $C$(H$\beta$) zero for 
the starburst knot.
In general, the derived line intensities for Tol 1214--277 are 
in fair agreement with those by Campbell et al. (1986) 
and Pagel et al. (1992) with the exception of [O\,III] $\lambda$4363 
and He II $\lambda$4686 which are stronger in our spectrum. 
Inspection of the higher resolution but lower 
signal-to-noise ratio spectrum of Pagel et al. (1992) shows 
the [Fe\,V] $\lambda$4227 line to be also present, 
although it was not discussed by those authors.

%
\subsubsection{Heavy element abundances}

\begin{figure*}[tbh]
\figurenum{4}
\plotfiddle{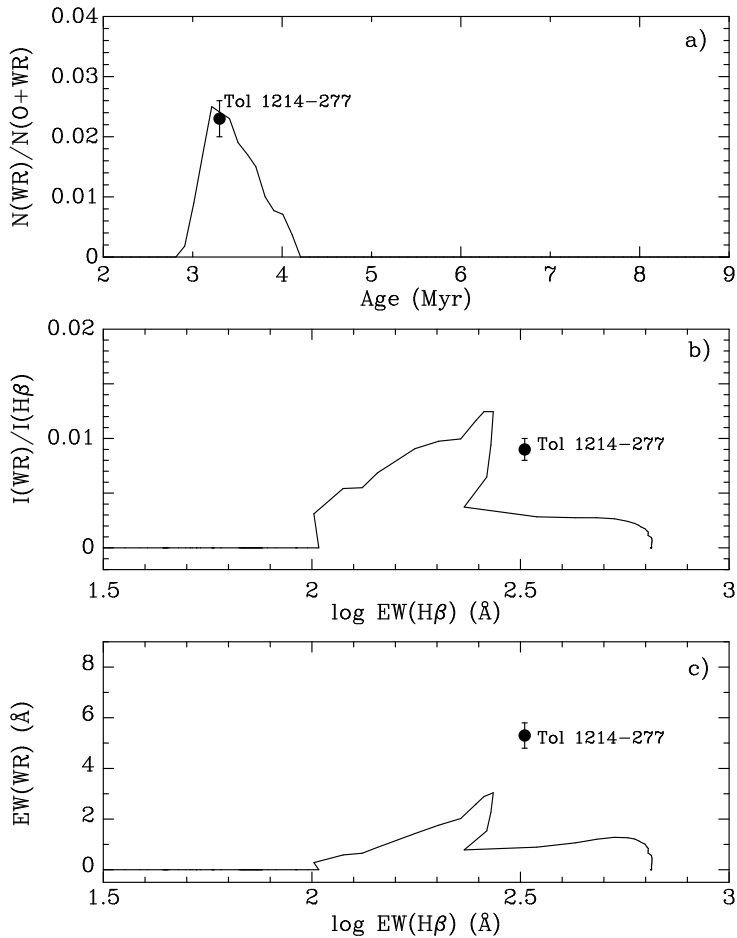}{0.cm}{0.}{60.}{60.}{-200.}{-400.}
\vspace{12.cm}
\figcaption{\label{f4} (a) The relative number of WR stars vs. age of an
instantaneous burst. (b) and (c) The relative flux and the equivalent width 
of the Wolf--Rayet blue bump vs. the equivalent width of the H$\beta$ emission 
line. The solid lines illustrate theoretical predictions from 
Schaerer \& Vacca (1998) for a heavy element mass fraction $Z$ = 0.001.}
\end{figure*}

The high signal-to-noise ratio VLT spectrum 
permits to derive element abundances with a higher precision 
than in previous studies.
However, the precision of such determinations is limited by the
absence of the VLT standard star observations as it is discussed in the
Sect. 2.2.
To derive heavy element abundances, 
we have followed the procedure detailed in 
Izotov, Thuan \& Lipovetsky (1994, 1997c, hereafter ITL94 and ITL97). 

We adopted a two-zone photoionized H\,II 
region model (Stasi\'nska 1990) including a high-ionization zone 
with temperature $T_e$(O\,III), and a low-ionization zone with 
temperature $T_e$(O II).  
We have determined $T_e$(O III) from the 
[O III]$\lambda$4363/($\lambda$4959+$\lambda$5007) ratio 
using a five-level atom model. That temperature is used for the 
derivation of the O$^{+2}$, Ne$^{+2}$ and Ar$^{+3}$ ionic 
abundances. 
To derive $T_e$(O II), we have utilized the relation between
$T_e$(O II) and $T_e$(O III) (ITL94), based on a fit to the
photoionization models of Stasi\'nska (1990). The temperature $T_e$(O II) is 
used to derive the O$^+$, N$^+$, S$^+$ 
and Fe$^+$ ion abundances. For Ar$^{+2}$
and S$^{+2}$ we have adopted an electron temperature intermediate between
$T_e$(O III) and $T_e$(O II) following the prescriptions of Garnett (1992).
The electron number density $N_e$(S II) (Table \ref{Fricke.tab3}) is derived from the [S II] 
$\lambda$6717/$\lambda$6731 ratio. 
The electron temperature $T_e$(O III) derived for Tol 1214--277 in this 
paper is higher than that in Campbell et al. (1986) and Pagel et al. (1992) 
because of
 the larger intensity of the [O\,III] $\lambda$4363 emission line.
The electron number density $N_e$(S II) is essentially the same as that derived
by Pagel et al. (1992). 

The oxygen abundance is derived as
\begin{equation}
{\rm \frac{O}{H} = \frac{O^+}{H^+} + \frac{O^{+2}}{H^+} + \frac{O^{+3}}{H^+}},
\label{eq:O} 
\end{equation}
where
\begin{equation}
{\rm \frac{O^{+3}}{O^+ + O^{+2}} = \frac{He^{+2}}{He^+}}.
\end{equation}

Total abundances of other heavy elements were computed after correction 
for unseen stages
of ionization as described in ITL94 and Thuan, Izotov \& Lipovetsky (1995).

The heavy element abundances obtained in this study
are in general agreement with those derived by Pagel et al. (1992). 
Our value for the oxygen abundance 12 + log (O/H) = 7.52 $\pm$ 0.01 
agrees well with the value 7.59 $\pm$ 0.05 reported by Pagel et al. (1992). 
This is also the case for the sulfur abundance as well as for  
the nitrogen-to-oxygen abundance ratio of log N/O = --1.50 $\pm$ 0.02 which
compares well with the value of --1.46 $\pm$ 0.06 obtained by Pagel et
al. (1992). 

%
\subsubsection{Helium abundance}
%
He emission-line strengths are converted to singly ionized helium 
$y^+$ $\equiv$ He$^+$/H$^+$ and doubly ionized helium $y^{+2}$ $\equiv$ 
He$^{+2}$/H$^+$
abundances using the theoretical He I recombination line emissivities 
by Smits (1996). 

To obtain the total helium abundance, the fraction of unseen neutral helium 
needs to be considered. 
In order to estimate the contribution of neutral helium, we have  
used the ``radiation softness parameter'' $\eta$ of V\'{\i}lchez \& Pagel (1988)
\begin{equation}
\eta = \frac{{\rm O}^+}{{\rm S}^+}\frac{{\rm S}^{+2}}{{\rm O}^{+2}}.       
\label{eq:eta}
\end{equation}
The fraction of neutral helium becomes significant ($\geq$ 5\%) when 
$\eta \geq$ 10 (Pagel et al. 1992). Given, however, that in \tol\ $\eta$ is 
equal to 0.40, the contribution of neutral helium is expected to be
very small ($<$ 1\%). Recently Ballantyne, Ferland \& Martin (2000)
have shown that the helium ionization correction factor is negligible when
the [O III] $\lambda$5007/[O I] $\lambda$6300 ratio is greater than 300 and/or
[O III] $\lambda$5007/H$\beta$ is greater than 5. Since both of those
 conditions are
fulfilled for Tol 1214--277 (Table \ref{Fricke.tab2}), we did not 
correct for the contribution of neutral helium.

In  addition, a strong nebular He II $\lambda$4686 emission
line was detected in the spectrum of Tol 1214--277. Therefore, we
have added the abundance of doubly ionized helium $y^{+2}$ 
to $y^+$. The value of $y$$^{+2}$ is 6.6\% of $y^+$ 
in Tol 1214--277, significantly higher than in other low-metallicity BCDs.

Finally the helium mass fraction was calculated as
\begin{equation}
Y=\frac{4y[1-20({\rm O/H})]}{1+4y},                     \label{eq:Y}
\end{equation}
where $y$ = $y^+$ + $y^{+2}$ is the number density of helium relative to 
hydrogen (Pagel et al. 1992).

The main mechanisms causing deviations of the He I emission line intensities 
from the recombination theory are collisional and fluorescent enhancement.     
In order to correct for these effects, we have adopted the following 
procedure,  
discussed in more detail in ITL94 and ITL97: using the formulae by Kingdon \& 
Ferland (1995) for collisional enhancement and the Izotov \& Thuan 
(1998) fits to Robbins (1968) calculations for fluorescent enhancement, we 
have evaluated the electron number density $N_e$(He II) and the optical depth 
$\tau$($\lambda$3889) in the He I $\lambda$3889 line in a self-consistent 
way, so that the He I $\lambda$3889/$\lambda$5876, 
$\lambda$4471/$\lambda$5876, 
$\lambda$6678/$\lambda$5876 and $\lambda$7065/$\lambda$5876 line ratios 
have their recombination values, after correction for collisional and 
fluorescent enhancement. 
Since the He I $\lambda$3889 line is blended with the H8 $\lambda$3889 line, we have 
subtracted the latter, assuming its intensity to be equal to 0.106 
$I$(H$\beta$) (Aller 1984).
The singly ionized helium abundance $y^+$ 
and He mass fraction $Y$ is obtained for each of the 
three He I $\lambda$4471, $\lambda$5876 and $\lambda$6678 lines
by the above mentioned self-consistent procedure. 
We then derived the weighted mean $y^+$ of these three determinations, the
weight of each line being scaled to its intensity. 
Note that this weighted mean may be an underestimate as
$y^+$($\lambda$4471) is lower than $y^+$($\lambda$5876) $\approx$
$y^+$($\lambda$6678) (cf. Table 3), possibly because of underlying 
stellar absorption being most important for the He I $\lambda$4471 emission line. 
%
\subsection{The Wolf-Rayet stellar population}
%


\begin{deluxetable}{lc}
\tablenum{4}
\tablecolumns{2}
\tablewidth{250pt}
\tablecaption{Parameters of the WR population
\label{Fricke.tab4}}
\tablehead{
\colhead{Parameter}
&\colhead{Tol 1214--277}}
\startdata
$k_{\rm corr}$\tablenotemark{a} &1.86 \nl
$F$(H$\beta$)$_{\rm corr}$\tablenotemark{b,c}
         &290$\pm$2 \nl
$F$(WR)\tablenotemark{c}
         &2.70$\pm$0.25 \nl
$EW$(WR) \AA
         &5.6$\pm$0.5 \nl
$N$(WR)
         &170$\pm$20 \nl 
$\eta_0$\tablenotemark{d}
         &1.2 \nl 
$N$(O)
         &6350$\pm$50 \nl 
$N$(WR) / ($N$(O + WR)
         &0.023$\pm$0.003 \nl 
\enddata
\tablenotetext{a}{aperture correction factor.}
\tablenotetext{b}{aperture-corrected H$\beta$ flux.}
\tablenotetext{c}{in units of 10$^{-16}$ ergs s$^{-1}$cm$^{-2}$.}
\tablenotetext{d}{Schaerer \& Vacca (1998).}
\end{deluxetable}

The flux and equivalent width of the blue bump 
in Tol 1214--277 measured after subtraction
of nebular emission are (2.70 $\pm$ 0.42) $\times$ 
10$^{-16}$ ergs s$^{-1}$cm$^{-2}$ and 5.6 \AA, respectively.
From the luminosity of the bump of
$L_{\rm WR}$ = 3.42 $\times$ 10$^{38}$ ergs s$^{-1}$ and assuming a 
luminosity of a single WNL star of 2.0 $\times$ 10$^{36}$ ergs s$^{-1}$ 
(Schaerer \& Vacca 1998) in the blue bump 
we estimate the number of WR stars in \tol\ 
to be $N_{\rm WR}$ = 170. We note however that $N_{\rm WR}$ is subject to
several uncertainties caused by the weakness of the bump, its contamination by
the emission of the nebular lines, the uncertainties in the adopted luminosity
of a single WR star in the blue bump and others. 
New observations with higher spectral resolution
and higher signal-to-noise ratio are desirable
to better constrain the spectral type of the WR stars and to
improve on the determination of $N_{\rm WR}$.

The number of O stars is deduced from the H$\beta$ luminosity
after subtracting the contribution of WR stars from it.
Because the H$\beta$ emission is extended and the slit does not 
cover the whole region of ionized gas emission, care should be 
exercised in correcting for aperture effects.
To estimate the fraction of the H$\beta$ flux escaping detection 
we followed the procedure developed by Guseva, Izotov \& Thuan (2000). 
Applying the correction factor of 1.86 obtained by that method, we 
infer the luminosity of the H$\beta$ emission line 
to be $L_{\rm cor}$(H$\beta$) = 3.68 $\times$ 10$^{40}$ ergs s$^{-1}$.

The number of O stars can be derived from the number of ionizing photons 
$Q_0^{\rm cor}$ which is related to the total luminosity of the H$\beta$ emission
line $L_{\rm cor}$(H$\beta$) by
\begin{equation}
L_{\rm cor}({\rm H}\beta) = 4.76\times10^{-13}Q^{\rm cor}_0.
\end{equation}
For a representative O7V star we adopt the
number of Lyman continuum photons emitted to be
$Q_0^{{\rm O7V}}$ = 1 $\times$ 10$^{49}$ s$^{-1}$ (Leitherer 1990).
The total number of O stars is then derived from the number of O7V stars by
correcting for other O stars subtypes, using the parameter $\eta_0$
introduced by Vacca \& Conti (1992) and Vacca (1994).
The quantity $\eta_0$ depends on the initial mass function 
for massive stars and is, in general, a function of time
because of their secular evolution (Schaerer 1996). 
Schaerer \& Vacca (1998) have calculated $\eta_0$ as a function of the
time elapsed from the onset of an instantaneous burst as  
inferred from the equivalent width $EW$(H$\beta$).
Adopting an IMF with a Salpeter slope $\alpha$ = 2.35 and lower
 and upper mass
limits of 0.8 $M_\odot$ and 120 $M_\odot$, we estimate in \tol\ from 
$EW$(H$\beta$) $\sim$ 320 \AA\ a burst age of 3.3 Myr and $\eta_0$$(t)$ = 1.2.

In the spectrum of Tol 1214--277 (Fig. \ref{f3}) the broad 
blue WR bump at $\lambda$4650 is detected.
On the other hand, no appreciable 
C\,IV $\lambda$4658, 5808 WR lines are found suggesting that 
the WR stellar population consists of late WN stars only.
This makes Tol 1214--277 the third known WR galaxy with oxygen abundance less
than 1/20 solar, besides
 I Zw 18 (Izotov et al. 1997a; Legrand et al. 1997) and
SBS 0335--052 (Izotov et al. 1999).

\begin{figure*}[tbh]
\figurenum{5}
\plotfiddle{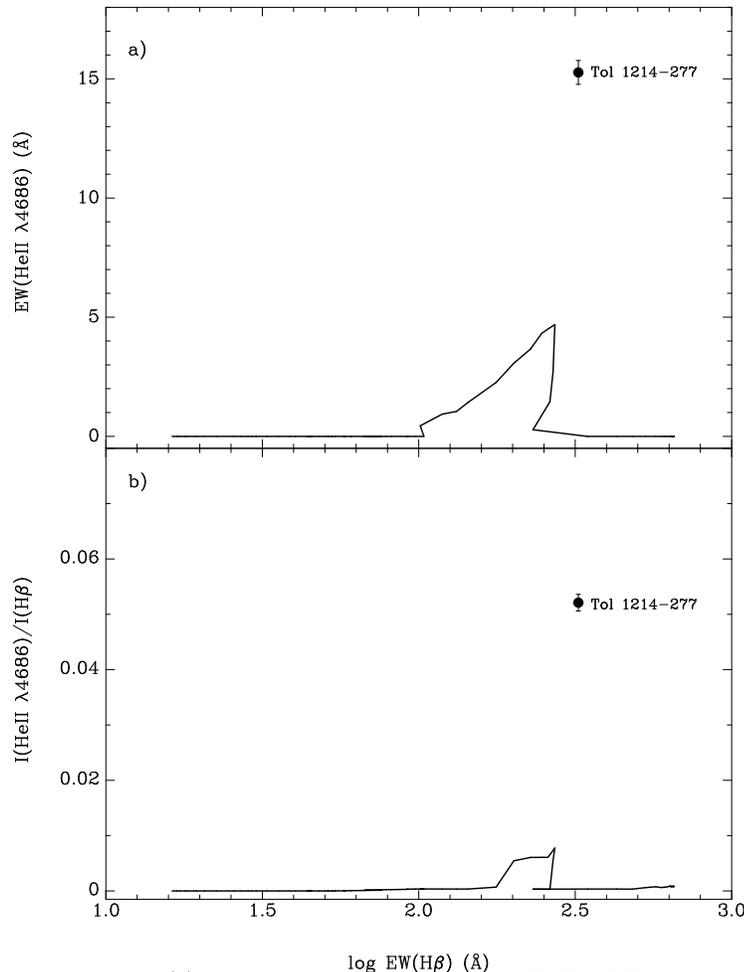}{0.cm}{0.}{50.}{50.}{-180.}{-380.}
\vspace{12.5cm}
\figcaption{\label{f5} (a) The equivalent width and (b) the intensity
ratio of the nebular He II $\lambda$4686 emission line and H$\beta$ 
vs. the H$\beta$ equivalent width in Tol 1214--277. 
Solid lines show theoretical predictions from Schaerer \&
Vacca (1998) for the heavy element mass fraction $Z$ = 0.001.}
\end{figure*}

\begin{figure*}[tbh]
\figurenum{6}
\plotfiddle{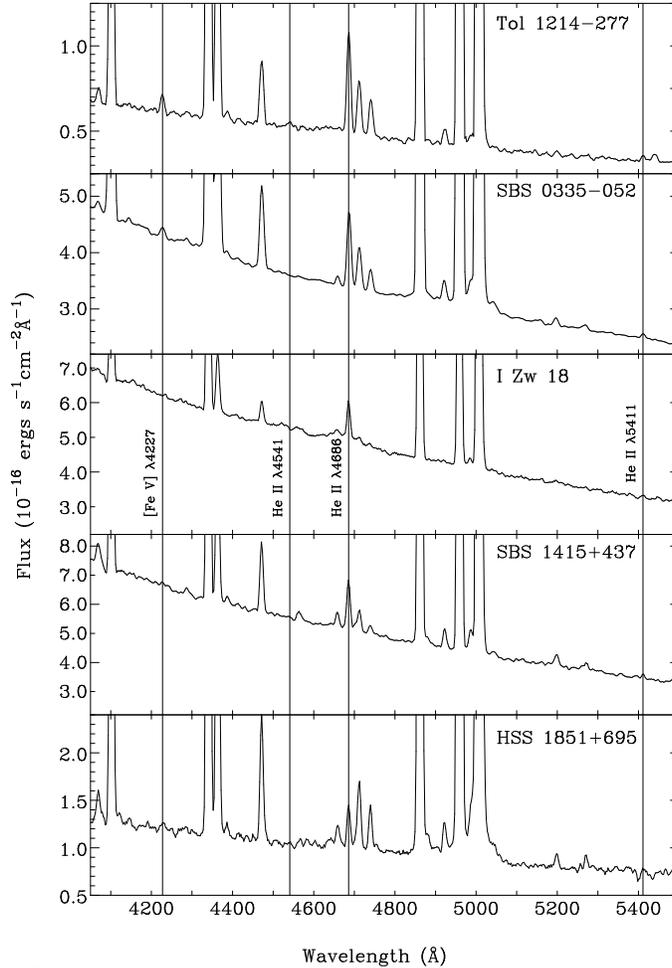}{0.cm}{0.}{50.}{50.}{-200.}{-380.}
\vspace{12.5cm}
\figcaption{\label{f6} The spectra of 5 low-metallicity blue compact dwarf galaxies with
detected nebular He II lines and the [Fe V]$\lambda$4227 line. The location
of lines is marked by vertical lines.}
\end{figure*}

To derive the number of O stars, it is 
necessary to subtract the contribution of WR stars from the total
number of ionizing photons. Following Schaerer, Contini \& Kunth (1999), we
 assume that 
the average Lyman continuum flux per WR star $Q^{\rm WR}_0$
is comparable to $Q_0^{\rm O7V}$ and equal to 
1.0 $\times$ $10^{49}$ s$^{-1}$. Thus: 

\begin{equation}
N({\rm O})=\frac{Q^{\rm cor}_0-N_{\rm WR}Q^{\rm WR}_0}{\eta_0(t)Q^{\rm O7V}_0}.
\label{eq:NO}
\end{equation}
From Eq.\,\ref{eq:NO}, the total number of O stars in Tol 1214--277 
is found to be $N$(O) = 6350 yielding a relative number of WR stars
$N$(WR) / $N$(O+WR) = 0.023. We point out here that $N$(O) 
and $N$(WR) / $N$(O+WR) should be considered, respectively, as a lower and 
upper limits because some part of Lyman continuum photons can escape the
H II region or will be absorbed by dust grains.

In Fig. \ref{f4} we compare the relative number of WR stars, relative flux
and equivalent width of the blue bump emission
with theoretical predictions by Schaerer \& Vacca (1998). 
The solid lines show model predictions for a heavy element
mass fraction $Z$ = 0.001. The relative number of WR stars in \tol\
is in good agreement with the theoretical value predicted for an 
instantaneous burst. 
The agreement is not so good, however, for the relative 
flux and equivalent width of the blue bump.
Both observed values correspond to larger $EW$(H$\beta$) than predicted 
by theory, i.e. to an earlier stage of an instantaneous burst. 
Furthermore, while the observed relative flux of the blue bump is smaller 
than the maximum theoretical value, its equivalent width 
is markedly larger than theoretical predictions.
This is also the case for the equivalent width and relative intensity of 
the nebular He\ II $\lambda$4686 emission line (Fig. \ref{f5}). 

%
\subsection{High-ionization emission lines}
%

Previous spectroscopic studies (e.g., Campbell et al. 1986)
have revealed that the hardness of the stellar ionizing radiation in BCDs
increases with decreasing metallicity. This trend implies that some nebular
emission lines of ions with high ionization potentials may be present in the
spectra of very metal-deficient BCDs such as Tol 1214--277. 
Indeed, the high signal-to-noise VLT spectrum of this galaxy 
allows for the detection of such lines. In particular, a strong 
nebular He II $\lambda$4686 emission line is observed in the spectrum 
of the brightest knot of Tol 1214--277 (Fig. \ref{f3}).
The presence of this emission implies that the 
hard radiation beyond the wavelength of 228 \AA\, equivalent 
to the ionization potential of 4 Ryd for He$^+$ ion, is strong. 

A strong nebular He II $\lambda$4686 emission line has also been detected
in many low-metallicity blue compact dwarf galaxies (in roughly 50\% of 
the samples investigated by ITL94, ITL97, Thuan et al. (1995, 1999), Izotov \&
Thuan (1998) and Izotov et al. (1996, 1997b)). 
Its intensity in some objects, including I Zw 18 and SBS 0335--052, exceeds 
$\sim$ 3\% of that of H$\beta$. The nebular He II emission in Tol 1214--277 has
been detected in previous studies by Campbell et al. (1986), Terlevich et al.
(1991), Pagel et al. (1992). Its intensity, however, has been inferred to be
$\sim$ 3\% of H$\beta$. Here we derive a significantly stronger 
He II $\lambda$4686 line exceeding 5\% of the intensity of the H$\beta$ 
emission line (Table \ref{Fricke.tab2}). This is the largest He II 
$\lambda$4686 / H$\beta$ ratio ever found in a BCD.

Furthermore, the high signal-to-noise ratio of the VLT spectrum has 
allowed to detect for the first time the weaker nebular He II $\lambda$4541 and 
$\lambda$5411 emission lines. Their intensities are in fair agreement with
the theoretical recombination values (Aller 1984) and the values measured in
some hot planetary nebulae (e.g., Feibelman et al. 1996; cf. 
Table \ref{Fricke.tab5}).

The origin of nebular He II $\lambda$4686 emission in photoionized supergiant
H II regions has been a subject of debate for years. 
The intensity of this line is several orders of magnitude larger 
than model predictions for photoionized H II regions 
(e.g. Stasi\'nska 1990). 

Schaerer (1996) synthesized the nebular and Wolf-Rayet He II $\lambda$4686 
emission in young starbursts. For heavy element mass fractions $Z_\odot$/5 
$\leq$ $Z$ $\leq$ $Z_\odot$, he predicted a strong nebular 
He II emission due to a significant fraction of WC stars in the early WR phases 
of the burst, and remarked that the predictions (typically 
$I$(He II) / $I$(H$\beta$) 
$\sim$ 0.01 -- 0.025) are in accord with observed values.
Schaerer \& Vacca (1998) proposed that hot WN stars may also play a
non-negligible role. 
Another mechanism, suggested by Garnett et al. (1991), is that radiative 
shocks in giant H II regions can produce relatively strong He II emission 
under certain conditions.

Further evidence for a hard UV radiation field in Tol 1214--277
comes from the detection of the strong nebular [Fe V] $\lambda$4227
emission line in its spectrum with an intensity of
$\sim$ 1\% of H$\beta$ (Table \ref{Fricke.tab2}).
In principle, in H II regions with strong He II emission the presence of
[Fe V] lines is not implausible since the ionization potential of the 
Fe$^{+3}$
ion is 4.028 Ryd, i.e. marginally higher than that of He$^+$. 
Therefore, it is expected that the locus of He$^{+2}$ emission 
is spatially associated with the Fe$^{+4}$ zone. 
Forbidden [Fe V] emission lines have already been seen in some hot planetary 
nebulae (e.g., Feibelman et al. 1996). To our knowledge, however, 
the observed [Fe V] $\lambda$4227 emission line in Tol 1214--277 is the 
first clear detection of spectral features of heavy element 
ions associated with the He$^{+2}$ zone in a BCD. 

\begin{figure*}[tbh]
\figurenum{7}
\plotfiddle{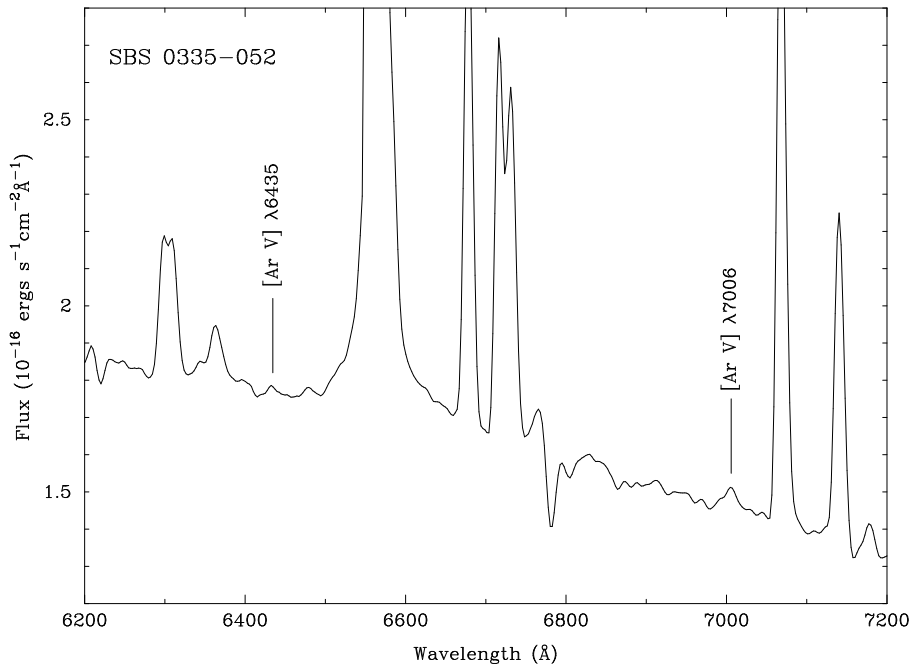}{0.cm}{270.}{50.}{50.}{-230.}{50.}
\vspace{8.5cm}
\figcaption{\label{f7} The spectrum of the red part of SBS 0335--052 with 
probable
nebular [Ar V] $\lambda$6435, $\lambda$7006 emission lines.}
\end{figure*}

We have checked for the possible presence of He II $\lambda$4541, 
$\lambda$5411 and [Fe V] $\lambda$4227 emission lines 
in spectra of other low metallicity BCDs using observations
by ITL94, ITL97, Izotov et al. (1996, 1997a, 1999), Thuan et al. (1999). 
Out of $\sim$ 50 galaxies, only in four other galaxies, mostly 
very metal-deficient, have some of these features been 
identified (Fig. \ref{f6}). 
In the spectrum of SBS 0335--052 (12 + log (O/H) = 7.30, Izotov et al. 1999) 
[Fe V] $\lambda$4227 and He II $\lambda$5411 emission lines are unambiguously
detected. Interestingly, in the red spectral region of SBS 0335--052, 
[Ar V] $\lambda$6435 and $\lambda$7006 emission lines are likely present
(Fig. \ref{f7}); they are absent, however, in spectra of other BCDs. 
The ionization potential of Ar$^{+3}$ of 4.396 Ryd is slightly larger than
that of He$^+$ and hence [Ar V] emission lines can be expected
in H II regions with strong nebular He II emission 
lines\footnote{Both [Ar V] lines in the spectrum of SBS 0335--052 are weak.
Therefore care must be exercised in their identification.
In addition, the [Ar V] $\lambda$7006 emission line can be
blended with other weak lines, particularly, with
the O I $\lambda$7002 line.}.
Possibly, a weak [Fe V] $\lambda$4227 emission line is present in the spectra
of SBS 1415+437 (12 + log(O/H) = 7.60, Thuan et al. 1999) and
HS 1851+695 (12 + log(O/H) = 7.78, Izotov et al. 1996), but it is absent in
the spectrum of the NW component of I Zw 18 (12 + log(O/H) = 7.13, Izotov et 
al. 1999). He II $\lambda$5411 is detected in I Zw 18 and SBS 1415+437.
All deep spectra shown in Fig. \ref{f6} have been obtained with 
large telescopes, such as Keck\,II\footnote{W.M. Keck Observatory 
is operated as a scientific partnership among the California Institute of 
Technology, the University of California and the National Aeronautics 
and Space Administration.  The Observatory was made possible by the 
generous financial support of the W.M. Keck Foundation.} 
(for SBS 0335--052) and the Multiple Mirror Telescope\footnote{Multiple 
Mirror Telescope Observatory is a joint facility of the Smithsonian
Institution and the University of Arizona.} (for other galaxies).
Therefore, the non-detection of the above mentioned relatively 
weak lines in previous studies of low-metallicity BCDs 
is probably due to an insufficient S/N ratio of those spectra.

\begin{figure*}[tbh]
\figurenum{8}
\plotfiddle{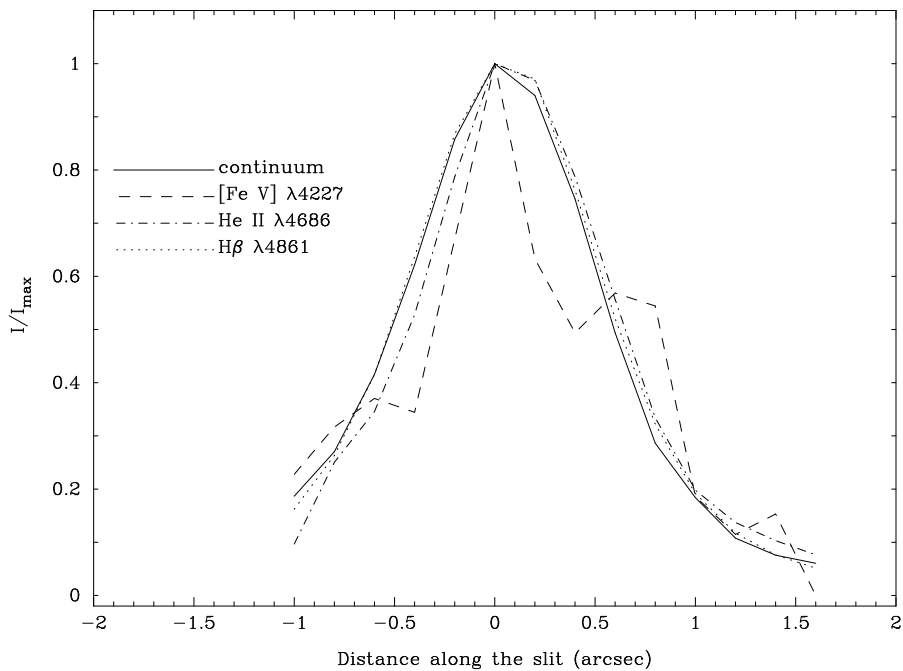}{0.cm}{270.}{50.}{50.}{-230.}{50.}
\vspace{8.5cm}
\figcaption{\label{f8} Distribution of the normalized intensity along
the slit for the H$\beta$ $\lambda$4861 emission line and for 
the adjacent continuum
as well as the He II $\lambda$4686 and [Fe V] $\lambda$4227 emission lines.}
\end{figure*}


\begin{deluxetable}{lcccccc}
\tablenum{5}
\tablecolumns{7}
\tablewidth{0pt}
\tablecaption{Relative intensities of high-ionization 
lines\label{Fricke.tab5}}
\tablehead{
\colhead{Parameter}
&\colhead{Tol 1214--277}&\colhead{SBS 0335--052}&\colhead{I Zw 18}
&\colhead{SBS 1415+437}&\colhead{IC 351}&\colhead{Theory\tablenotemark{a}}}
\startdata
$I$($\lambda$4541)/$I$($\lambda$4686)
         &0.043$\pm$0.027&  \nodata     &    \nodata   &   \nodata    & 0.033 & 0.035  \nl
$I$($\lambda$5411)/$I$($\lambda$4686)
         &0.068$\pm$0.020&0.065$\pm$0.014&0.085$\pm$0.044&0.098$\pm$0.036      & 0.085 & 0.081 \nl
$I$($\lambda$4227)/$I$($\lambda$4686)
         &0.181$\pm$0.025&0.178$\pm$0.022&    \nodata   &   \nodata   & 0.009 &\nodata \nl
$I$($\lambda$6435)/$I$($\lambda$4686)
         &    \nodata   &0.018$\pm$0.012&    \nodata   &    \nodata   & 0.005 &\nodata \nl
$I$($\lambda$7005)/$I$($\lambda$4686)
         &    \nodata   &0.035$\pm$0.019&    \nodata   &    \nodata   & 0.009 &\nodata \nl
\enddata
\tablenotetext{a}{Aller (1984).}
\end{deluxetable}

The [Fe V] $\lambda$4227 and He II $\lambda$4686 intensity distribution 
along the slit in Tol 1214--277 is shown in Fig. \ref{f8} along
with the distribution of H$\beta$ and the continuum intensity. 
The spatial profile of the He II $\lambda$4686 emission line has 
a FWHM $\sim$ 1\arcsec\ which is slightly higher than the
seeing during the observations. The intensity distribution 
of the [Fe V] $\lambda$4227 emission line along the slit seems 
narrower than that of the He II $\lambda$4686 emission line.
However, the difference is not significant 
because of the low intensity of the [Fe V] $\lambda$4227 emission line.
The small angular size of the bright H II region where high ionization
emission lines are observed precludes a comparative study of the
spatial distribution of high and low ionization species in the starburst 
region of \tol. From Fig. \ref{f8} we conclude that the zones of 
He II $\lambda$4686 and H$\beta$ emission are most likely spatially 
coincident. 
A coincidence of the He$^{+2}$ and H$^+$ zones has
also been seen for the NW component of I Zw 18 (e.g., Izotov et al. 1999). 
On the other hand, in SBS 0335--052 the He$^{+2}$ zone is found to be 
offset relative to the H$^{+}$ zone toward the evolved stellar clusters 
(Izotov et al. 1997b) suggesting that the presence of
He$^{+2}$ is likely related to the ionizing radiation of post-main-sequence 
stars or radiative shocks produced by supernovae (SNe). 

Some support for the hypothesis that high ionization 
species are produced by SN shocks comes from the comparison 
of the intensities of the [Fe V] $\lambda$4227, 
and probable [Ar V] $\lambda$6435, $\lambda$7006
emission lines relative to the intensity of He II $\lambda$4686
in the most metal-deficient BCDs and planetary nebulae. 
From Table \ref{Fricke.tab5}, it is evident that 
the $I$([Fe V] $\lambda$4227) / $I$(He II $\lambda$4686) 
ratio in low-metallicity BCDs is more than one order of magnitude larger 
than that in the planetary nebula IC 351 (Feibelman et al. 1996). 
Furthermore, the relative intensities of [Ar V] emission lines 
in SBS 0335--052 
are $\sim$ 3 -- 4 times larger than those in the same planetary nebula.
However, the latter estimate is uncertain due to the weakness of the
[Ar V] emission lines.
The electron temperature in the shocked material is expected to be higher 
than that in the rest of the H II region. Collisionally excited forbidden 
lines are more sensitive to the electron temperature as compared to the 
recombination He II emission lines. Therefore, at high temperatures, 
produced by fast shocks, we expect larger intensities of [Fe V] and [Ar V]
emission lines relative to He II emission lines compared to those in
planetary nebulae.

\begin{figure*}[tbh]
\figurenum{9}
\plotfiddle{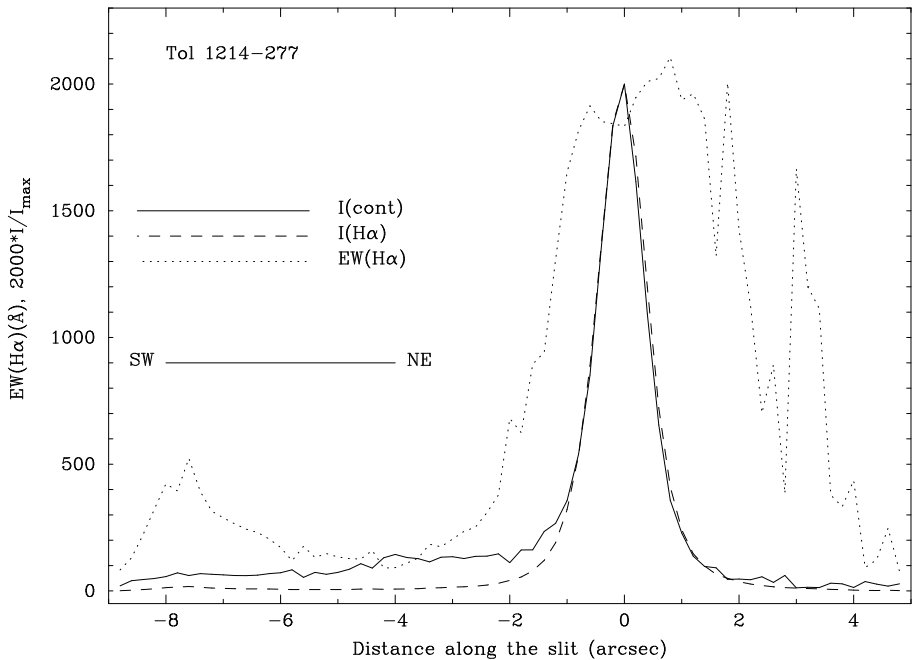}{0.cm}{270.}{50.}{50.}{-230.}{50.}
\vspace{8.5cm}
\figcaption{\label{f9} The distribution along the slit with P.A. = +39 deg 
of the H$\alpha$ equivalent width $EW$(H$\alpha$) and of the normalized intensities of the 
continuum and of the H$\alpha$ $\lambda$6563 emission line in Tol 1214--277.}
\end{figure*}

%
\section{Age of the underlying stellar population}
%
The large distance of \tol\ precludes the study of its
stellar composition by means of color-magnitude diagrams 
in the manner done for nearby BCDs 
(e.g. Schulte-Ladbeck et al. 1998, Lynds et al. 1998,
Aloisi et al. 1999, \"Ostlin 2000). 
Therefore, our analysis of the evolutionary status of 
\tol\ is based on the integrated colors and 
spectral energy distributions (SED) of its extended underlying emission.

The blue colors of the LSB host of \tol\ 
($U-B$) $\sim$ --0.4 mag and ($B-R$) $\sim$ +0.4 -- +0.5 mag
(Fig. \ref{f2}b) suggest that its faint underlying stellar component
extending to the SW is rather young.
However, care should be exercised in correcting the observed colors for 
gaseous emission which, judging from Fig. \ref{f3}, is 
important in the vicinity of the starburst knot.
To quantify the line-of-sight contribution of ionized gas to 
the emission of the underlying LSB galaxy we show in 
Fig. \ref{f9} the H$\alpha$ equivalent width $EW$(H$\alpha$) distribution
as well as that of the intensity of the H$\alpha$ line 
and of the adjacent continuum along the major axis of the BCD.
It is evident that while both intensity distributions are 
strongly peaked at  the starburst's location, 
the $EW$(H$\alpha$) distribution is broader and asymmetric.
This indicates that the gaseous emission dominates 
the line-of-sight emission to the NE from the 
starburst region, while it is not important in the SW direction
where the stellar background is strong.
Nevertheless, contamination by ionized gas cannot be 
entirely neglected even at intermediate radii, since one-dimensional
surface brightness profiles are derived averaging over the light from 
different regions with a varying degree of contribution 
by ionized gas. 
The fractional contribution of the ionized gas 
is expected to be larger in the $R$ band due to the H$\alpha$ line, 
a fact which may explain the slight reddening of the ($B-R$)
color by $\sim$ 0.1 mag at the effective radii between 4\arcsec\ and 5\arcsec\
(Fig. \ref{f2}b). 

\begin{figure*}[tbh]
\figurenum{10}
\plotfiddle{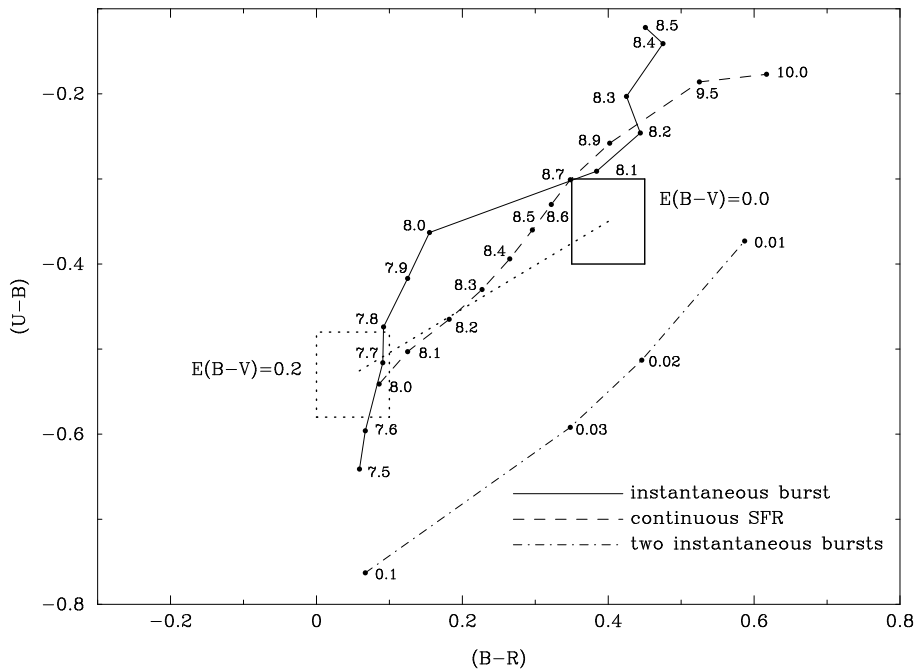}{0.cm}{270.}{50.}{50.}{-230.}{50.}
\vspace{8.5cm}
\figcaption{\label{f10} The $(U-B)$ vs. $(B-R)$ diagram for the underlying 
stellar component of Tol 1214--277. The solid box represents the observed 
value with an uncertainty of 0.05 mag while the dotted box indicates the position
of the LSB host of \tol\ when corrected for an extinction with $E(B-V)$ = 0.2 mag.
The solid line illustrates the evolution of an instantaneous burst
 as a function of log ($t$/yr). 
The dashed line connects model predictions for a continuous
constant star formation from 
the time log ($t$/yr) which labels each point to log($t$/yr) = 7.3.
The dot-dashed line connects values predicted for two instantaneous 
bursts with log ($t$/yr) = 7.3 and 9.3. Points are labelled with the
stellar mass fraction of the younger burst. All model predictions are computed 
using the Padua tracks with $Z$ = $Z_\odot$/20 (Bertelli et al. 1994).}
\end{figure*}

As characteristic colors of the  
stellar LSB component, we shall adopt ($U-B$) = --0.35 mag and 
($B-R$) = +0.4 mag with an uncertainty of 0.05 mag in either of them. 
These colors are shown in the 
($U-B$) vs. ($B-R$) diagram (Fig. \ref{f10}) by solid boxes. 
From the present spectral data, we have no information on 
the average extinction in the outskirts of \tol\ 
because of the faintness of the hydrogen emission lines. As an illustrative 
example, we
show with the dotted box the position of the underlying dwarf 
galaxy if a reddening $E$($B-V$) = 0.2 mag is adopted. 
Larger values for $E$($B-V$) are unlikely because in this
case the location of the underlying component in the ($U-B$) vs. ($B-R$) 
color diagram cannot be reproduced by any of the models.
Note that a correction for any residual contribution of gaseous 
emission along the major axis of the BCD will
shift the observed ($B-R$) index bluewards.
Thus we expect the colors of  
Tol 1214--277's underlying component to lie between 
the solid and dotted boxes in the ($U-B$) vs. ($B-R$) diagram
(Fig. \ref{f10}).
In the same figure we overlay model predictions for stellar 
populations with three different star formation histories:
the thin solid line illustrates the location expected for a stellar population 
formed in an instantaneous burst as function of log ($t$/yr); 
the dashed line connects model predictions for a continuous star 
formation at constant rate having started
at log ($t$/yr) as given by the label of each point and continuing 
to log ($t$/yr) = 7.3;
the dot-dashed line connects values predicted for two 
instantaneous bursts with ages log ($t$/yr) = 9.3 and 7.3.
In the last case, model predictions are labelled by the 
stellar mass fraction formed in the youngest burst.
Our models are computed using the stellar isochrones of 
Bertelli et al. (1994) and the compilation of stellar 
atmosphere models from Lejeune et al. (1994) for a metallicity 
$Z_{\odot}$/20.  
A Salpeter initial mass function with slope --2.35 and with lower and 
upper mass limits of 0.6 $M_{\odot}$ and 100 $M_{\odot}$ has been adopted.

It may be seen from Fig. \ref{f10} that the observed colors 
of the LSB host cannot be accounted for by the model with two 
instantaneous bursts, one of which is young and the other old. 
The observations can, however, be reproduced equally 
well by an instantaneously formed stellar population with an
age log ($t$/yr) $\la$ 8.1 and a stellar population forming continuously 
between log ($t$/yr) $\la$ 8.7 and log ($t$/yr) $=$ 7.3.

\begin{figure*}[tbh]
\figurenum{11}
\plotfiddle{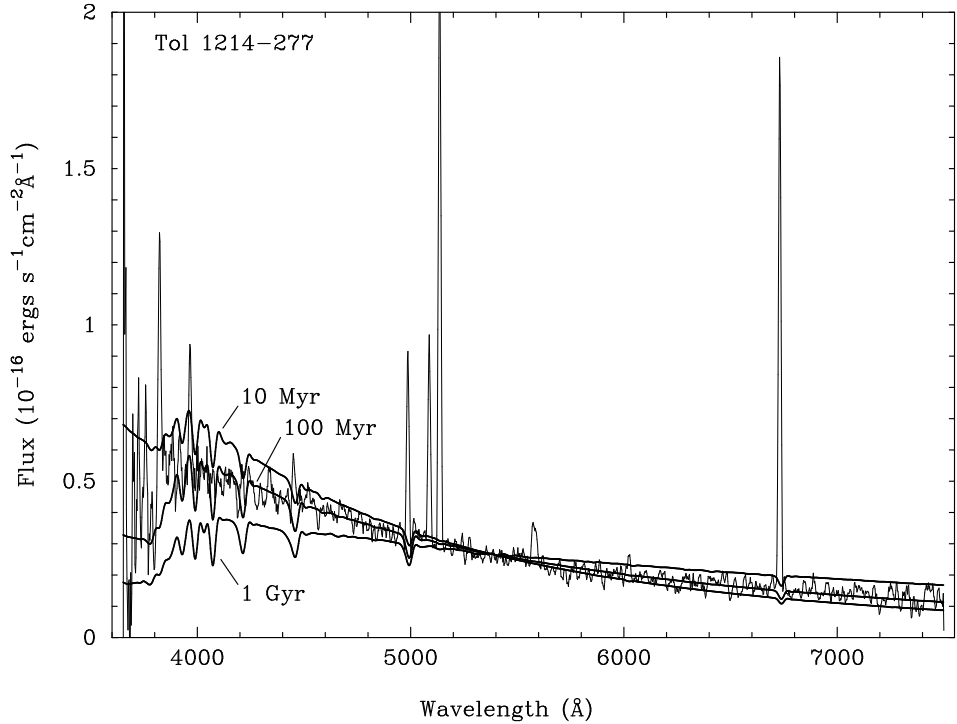}{0.cm}{270.}{60.}{60.}{-260.}{70.}
\vspace{9.5cm}
\figcaption{\label{f11} 
Spectrum of the underlying LSB host of \tol\ on which are superposed the theoretical
spectral energy distributions (SED) of stellar populations with ages 10 Myr, 100 Myr and 1
Gyr. The theoretical SEDs are calculated using isochrones from Bertelli et
al. (1994) and the stellar atmosphere model compilation from Lejeune et
al. (1998), for a metallicity $Z_{\odot}$/20. 
The best fit is achieved with a theoretical SED with age $\sim 100$ Myr.}
\end{figure*}

Further support in favour of the evolutionary youth of \tol\ is provided 
by the SED of its LSB host (Fig. \ref{f11}). Overlayed with the observed 
spectrum we show model SEDs for three instantaneous burst populations 
with ages 10 Myr, 100 Myr and 1 Gyr, computed with the models described above. 
It can be seen from Fig. \ref{f11} that the observed SED of the 
underlying LSB host of \tol\ is matched best with a single burst population 
with an age 100 Myr.

In summary, models with simple   
star formation histories of \tol\ give an age 
between 0.1 and 0.5 Gyr for its stellar underlying host.
Given that there is no compelling evidence for 
an appreciable stellar population with cosmological age,
we consider 1 Gyr to be a reasonable upper age limit for the BCD.
This conclusion is consistent with the age estimates based on the
color predictions from Geneva evolutionary tracks (Leitherer et al. 1999).

Therefore, \tol\ is a second example
besides SBS 1415+437 (Thuan et al. 1999) of an extremely 
metal-deficient BCD showing cometary morphology and
evidence for being relatively unevolved. 
This result supports the conjecture by Noeske et al. (2000) 
that cometary BCDs may be systematically younger than BCDs having a 
smooth elliptical or circular stellar LSB envelope (iE/nE BCDs).
Low metallicity may be another necessary condition for young age 
(Izotov \& Thuan 1999).
%
\section{Summary}
%
The main conclusions drawn from our imaging and spectroscopic analysis of
deep VLT data of the extremely metal-deficient ($Z$$\sim$$Z$$_{\odot}$/25) 
and nearby ($D$ = 103.9 Mpc) BCD \tol\ may be summarized as follows:

1. \tol\ undergoes a vigorous burst of star formation having 
ignited less than 4 Myr ago.
The starburst takes place within a bright ($M_B\!\sim\!-16$ mag) 
compact ($\la$ 500 pc in diameter) region,  giving rise to extended 
and abundant ionized gas emission with an H$\beta$ equivalent 
width of $\sim$ 320 \AA. 
The starburst is powered by several thousands of O7V stars and 170
late-type nitrogen Wolf--Rayet stars.

2. In this very metal-deficient BCD we discover for the first 
time the high ionization line [Fe V] $\lambda$4227. 
Moreover, we detect extraordinarily strong He II $\lambda$4686 
emission with an intensity as high as 5\% of that of the H$\beta$
emission line. This implies the presence of a very 
hard radiation field in \tol.
The intensity ratio $I$([Fe V] $\lambda$4227) / $I$(He II $\lambda$4686) 
in \tol\ compares well with that in another extremely metal-poor
BCD with [Fe V] $\lambda$4227 emission, SBS 0335--052, 
being in both cases larger by more than one order of magnitude than 
the ratio observed in high-excitation planetary nebulae. 
While the relative number of WR stars of $N$(WR) / $N$(O + WR) = 0.023 
in \tol\ is compatible with theoretical predictions,
the intensity of the He II $\lambda$4686 emission line exceeds
several times the predictions of standard H II photoionization 
models, even when the hard radiation component of Wolf-Rayet stars is 
taken into account. 
Therefore, we argue that the hard ionizing radiation field in \tol\
is produced from the combined effect of massive stars and SN-driven
shocks. 

3. Star-forming activity in \tol\ is confined to the northeastern tip 
of a cometary dwarf galaxy with an absolute $B$ magnitude $\ga$ --16 mag 
and an isophotal size of 7.6$\times$4.8 kpc at 28 $B$ \sbb.  
An exponential fitting law provides a reasonable approximation to
the intensity distribution of the stellar LSB host in its outskirts,
 for $\mu_B\ga 24.5$ \sbb. It fails, however, 
to properly describe the observed brightness distribution 
at intermediate and high intensity levels. These are better fitted 
by an exponential distribution which flattens at small radii,
similar to the V-type profiles described by Binggeli \& Cameron (1991).

4. The radially averaged ($U-B$) and ($B-R$) colors of the LSB host
of \tol\ are consistent with those for
either an instantaneous burst 
with log ($t$/yr) $\la$ 8.1, or a continuous star formation 
between log ($t$/yr) $\la$ 8.7 and log ($t$/yr) $=$ 7.3. 
We however cannot definitely exclude the presence of a small fraction of 
old (age $>$ 1 Gyr) stars due to their intrinsic faintness.
%


\acknowledgements
Research by K.J.F. and P.P. has been supported by the
Deutsches Zentrum f\"{u}r Luft-- und Raumfahrt e.V. (DLR) under
grant 50\ OR\  9907\ 7. These authors, Y.I.I. and N.G.G. 
acknowledge support by the Volkswagen Foundation under grant No. I/72919.
Y.I.I. and T.X.T thank for partial financial support through NSF grant AST-9616863.
Y.I.I. and N.G.G. also thank for INTAS 97-0033 grant and for hospitality
at G\"ottingen Observatory. 
Y.I.I. thanks for a Gauss professorship of the G\"ottingen Academy of Sciences and
N.G.G. thanks for DFG grant 436 UKR 17/1/00.
All authors thank Drs. K.\ Reinsch (G\"ottingen) 
and S.\ Wagner (Heidelberg) for carrying out the GTO observing program.
Building the VLT spectrograph FORS at G\"ottingen was supported by BMBW/DESY under grant
053\ GO\ 10A.


{}

\end{document}